\documentclass[3p,times,twocolumn]{elsarticle}

\usepackage{amssymb}

\begin{document}

\begin{frontmatter}




\title{Third Quantization and Quantum Universes}


\author{Sang Pyo Kim}

\ead{sangkim@kunsan.ac.kr}

\address{Department of Physics, Kunsan National University, Kunsan 573-701, Korea}

\begin{abstract}
We study the third quantization of the Friedmann-Robertson-Walker cosmology
with $N$-minimal massless fields.
The third quantized Hamiltonian for the WDW equation in the minisuperspace
consists of infinite number of intrinsic time-dependent, decoupled oscillators.
The Hamiltonian has a pair of invariant operators for each universe
with conserved momenta of the fields that play a role of the
annihilation and the creation operators
and that construct various quantum states for the universe. The closed universe
exhibits an interesting feature of transitions from stable states to tachyonic states
depending on the conserved momenta of the fields. In the classical forbidden unstable regime,
the quantum states have googolplex growing position
and conjugate momentum dispersions,
which defy any measurements of the position of the universe.
\end{abstract}


\end{frontmatter}


\section{Introduction}

The recent seven-year Wilkinson Microwave Anisotropy Probe (WMAP) observation \cite{WMAP7}
and South Pole Telescope Sunyaev-Zel'dovich (SPT-SZ)
survey \cite{SPT} support the $\Lambda$CDM model with an inflaton.
The spacetime of the universe is the Friedmann-Robertson-Walker (FRW) geometry
with a cosmological constant
and the matter field is a single large-field inflaton with a power not greater than two.
It would thus be interesting to study the FRW universe with a scalar field beyond the
inflationary era. The quantum fluctuations of
not only matter fields but also spacetimes become important
far beyond the inflation period toward the Big Bang singularity.

Hawking and Penrose have shown that under the weak energy condition
all timelike (null) geodesics end at a singularity and the universe must have
the Big Bang singularity \cite{hawking-penrose}. In many inflationary models
the weak energy condition is violated but under the averaged Hubble expansion
the universe has also a beginning \cite{BGV}. General relativity necessarily
ends up with with the Big Bang singularity, the classical theory breaks down and
geometries and matters suffer quantum fluctuations.
Therefore understanding the universe including the Big Bang era
necessarily requires quantum gravity.

The quantum geometrodynamics based on the Wheeler-DeWitt (WDW) equation is a canonical approach,
in which three-geometries of spacelike hypersurfaces and their extrinsic curvatures
are conjugate pairs \cite{dewitt}. The advantage of the geometrodynamic
approach is that the variables present in
the Einstein equation for cosmological models are quantized
and that semiclassical gravity emerges from oscillatory wave packets
along classical spacetime trajectories but still keeping quantum fluctuations of matters.
Another quantum gravity is the path integral over all geometries and matters,
which is substantiated by Hartle-Hawking's no-boundary wave function
that sums all compact Euclidean geometries which match with the Lorentzian
geometry on a three-hypersurface \cite{hartle-hawking,hawking84}.

The Hartle-Hawking wave function seems to be qualitatively favored
by the current cosmological observations \cite{page03}.
However, it would be interesting to revisit quantitatively quantum FRW cosmology with minimal scalar fields with monomial and general potentials \cite{kim91,kim92,kim-page92}.
One advantage of quantum cosmology based on the WDW equation
is that semiclassical gravity emerges from WDW equation
along wave packets and recovers unitary quantum field theory in the curved spacetime.
Thus it provides a natural and consistent framework to study the back-reaction problem \cite{kim95,kim96,BFV,kim97}.
For instance, quantum FRW cosmology with massless
fields together with a cosmological constant explains the back-reaction of de Sitter
radiation \cite{GKS}, which is equivalent to the back-reaction of Hawking radiation
in a de Sitter Schwarzschild black hole \cite{GPS}.

This paper is the first of series of works on the third quantized formulation of quantum FRW cosmology with a minimal scalar field
\begin{eqnarray}
\bigl[ - \frac{\partial^2}{ \partial \alpha^2} +
\frac{\partial^2}{\partial \phi^2} + V_m(\phi, \alpha) + V_g(\alpha) \Bigr] \Psi (\alpha, \phi) = 0 \label{FRW scalar}
\end{eqnarray}
with $e^{\alpha}$ being the scale factor and $V_g$ denoting the terms from the three-geometry
curvature and the cosmological constant of the FRW geometry.
Quantum cosmology with a minimal massive scalar field was studied by Hawking \cite{hawking84} and
with a minimal massive field in perturbed FRW geometry by Halliwell and Hawking \cite{halliwell-hawking85}. Quantum FRW cosmology (\ref{FRW scalar}) with minimal scalar fields with monomial and general potentials
was intensively studied by Kim \cite{kim91,kim92} and Kim and Page \cite{kim-page92}.
A remarkable prediction of the quantum FRW cosmology with a minimal scalar field
is that the wave functions (\ref{FRW scalar}) may not be singular even at the Big Bang singularity \cite{kim92}.
The set of wave functions of the WDW equation is infinite corresponding to infinitely many
Cauchy initial data \cite{kim92,mostafazadeh98}.

Each wave function of the WDW equation (\ref{FRW scalar})
describes a universe with fixed physical parameters
such as the cosmological constant, the topology of the geometry and quanta for scalar fields.
The proper context to interpret the wave functions may associate them with the
operators that create or annihilate the wave functions. The quantization of the WDW equation
is called the third quantization or quantum field theory since the WDW equation involves
the second quantized matters. In the third quantization
the WDW equation is a wave operator
residing on the superspace and the wave functions are associated with the corresponding
creation and annihilation operators
\cite{mcguigan88,hosoya-morikawa89,mcguigan89,mcguigan91,peleg91,zhuk92,KKS,abe93,horiguchi93,CGMT,BGMU,DGOV,perez-diaz10a,perez-diaz10b,perez-diaz12,bousso-susskind12}. Thus the Hilbert space consists of the vacuum of no-universe,
one-universe state and multi-universe states.

In the third quantization for quantum cosmology,
Hosoya and Morikawa \cite{hosoya-morikawa89} studied
a spatially flat FRW cosmology minimally coupled to a massless scalar field
and Abe \cite{abe93} and Horiguchi \cite{horiguchi93}
also studied an open universe with a cosmological constant
with or without a massless field. In this paper we shall study quantum cosmology
for an open or closed FRW universe with $N$-massless scalar fields. The wave function of the WDW equation could be found in terms of the Bessel and the Hankel functions for the open universe
and in terms of the Bessel and the modified Bessel functions for the closed universe.
The compact spaces with constant curvature in the multidimenional
quantum cosmological model play the same role of the massless fields
in the WDW equation \cite{zhuk92,IMZ}.
A spatially flat universe will not be considered since the WDW equation is the same a free Klein-Gordon equation. Each wave function for given momenta of the massless fields carries the evolution of the universe in the third quantized formulation. The three-curvature of the FRW geometry provides an intrinsic time-dependent term. This Hamiltonian for the WDW equation is equivalent
to an infinite number of decoupled time-dependent oscillators, whose
quantum states could be found in terms of the invariant operators that play the role of the time-dependent annihilation and creation operators. These operators for each universe with the given momenta and with the given topology construct interesting wave functions, such as the Gaussian wave functions, the number-state wave functions and the coherent states.

The organization of this paper is as follows. In Sec. \ref{sec-II} we review the WDW equation
for the FRW universe minimally coupled to $N$-massless scalar fields and find the solutions
for the open and closed universes. In Sec. \ref{sec-III} we formulate the quantum FRW cosmology
in the third quantization and introduce pair of invariant operators that act as the time-dependent
annihilation and creation operators for each momenta of the fields. In particular, the
Gaussian wave functions that are dual to each other, not necessarily complex conjugate,
are introduced. The dispersion relation and the uncertainty are computed in terms of the solutions
for the WDW equation. In Sec. \ref{sec-IV} we discuss the production of baby universes during
the evolution of universe. The closed universe takes tachyonic states for
super-Planckian size, which is analogous to
the second order phase transition. We advance a method to quantize this tachyonic universe.

\section{Wheeler-DeWitt Equation} \label{sec-II}

We consider quantum cosmology for a FRW universe with $N$-minimal massless scalar fields. The
WDW equation takes the form (in the Planckian units of $\hbar = c = G = 1$)
\begin{eqnarray}
\bigl[ \square - {\cal M}^2 (\alpha) \Bigr] \Psi (\alpha, \phi_n) = 0, \label{wdw eq}
\end{eqnarray}
where $e^{\alpha} $ is the scale factor and $\square$ is the D'Alembertian operator
\begin{eqnarray}
\square = - \frac{\partial^2}{ \partial \alpha^2} + \sum_{n = 1}^{N}
\frac{\partial^2}{\partial \phi_n^2}, \label{dalem}
\end{eqnarray}
and the $\alpha$-dependent mass
\begin{eqnarray}
{\cal M}^2 (\alpha) = - k e^{4 \alpha} \label{mass}
\end{eqnarray}
comes from the scalar curvature of the three-geometry multiplied twice by the volume factor,
where $k = 1, 0$ and $-1$ are for a closed, flat and open universe, respectively.
The minisuperspace for the scale factor and the scalar fields has
the Lorentzian spacetime metric in $1+N$ dimensions
\begin{eqnarray}
ds^2 = - d \alpha^2 + \sum_{n=1}^{N} d \phi_n^2, \label{super sp}
\end{eqnarray}
on which eq. (\ref{dalem}) is a wave operator.
The multi-dimensional manifold $R \times M_1 \times \cdots \times M_N$ for each compact space
$M_n$ of constant curvature  \cite{zhuk92,IMZ} has the minisuperspace of the form (\ref{super sp}).

The fact that time does not appear in the super-Hamiltonian and the super-momenta constraints in
the ADM formalism raises the question of the origin of time in canonical quantum
gravity, i.e, the WDW equation. It should be reminded that the superspace for
a globally hyperbolic spacetime, such as the FRW geometry, has a Lorentzian signature
and one of the variables may play an intrinsic time in the superspace.
Then the WDW equation is the $(1+N)$-dimensional wave equation in the superspace with
a time-dependent mass. Hence it is legitimate to regard $\alpha$ as the intrinsic time.
Any function of $\alpha$ may play the same role, which corresponds to
a reparametrization of the wave operator. The choice of $\alpha$ makes the WDW equation
resembling the wave equation in the Minkowski spacetime with the time-dependent mass.

Each field $\phi_n$ has a conserved momentum $p_n$ related to $e^{i p_n \phi_n}$.
The solution of the form
\begin{eqnarray}
\Psi (\alpha, \vec{\phi}) = e^{ i \vec{p} \cdot \vec{\phi} } \psi_{\vec{p}} (\alpha),
\end{eqnarray}
with $\vec{\phi} = (\phi_1, \cdots, \phi_N)$ and $\vec{p} = (p_1, \cdots, p_N)$,
separates the WDW equation as
\begin{eqnarray}
\Bigl[\frac{\partial^2}{ \partial \alpha^2} + \vec{p}^2 - k e^{4 \alpha}
\Bigr] \psi_{\vec{p}} (\alpha) = 0. \label{grav eq}
\end{eqnarray}
The solutions to eq. (\ref{grav eq}) for $k = -1$ are given
by the Bessel and the Hankel functions \cite{table}
\begin{eqnarray}
\psi_{\vec{p}} (\alpha) = Z_{\pm ip/2} \Bigl(\frac{1}{2} e^{2 \alpha} \Bigr), \label{k=-1 sol}
\end{eqnarray}
and for $k =1$ by the Bessel functions and the modified Bessel functions
\begin{eqnarray}
\psi_{\vec{p}} (\alpha) = I_{\pm ip/2} \Bigl(\frac{1}{2} e^{2 \alpha} \Bigr), \quad K_{\pm ip/2} \Bigl(\frac{1}{2} e^{2 \alpha} \Bigr). \label{k=1 sol}
\end{eqnarray}
Note that the solutions (\ref{k=1 sol}) can be obtained by analytically continuing eq. (\ref{k=-1 sol})
through $\alpha + i \pi/4$, which changes the sign of $k$ in the equation (\ref{grav eq}). The case of $k = 0$
exactly takes the form of the massless Klein-Gordon equation.

There is another question of interpreting the wave functions for the WDW equation.
The probabilistic interpretation of the wave functions is problematic as
for the Klein-Gordon equation. However, with respect to the intrinsic time $\alpha$, the WDW equation may have the inner product \cite{KKS}
\begin{eqnarray}
\langle \Psi_{I} (\alpha, \vec{\phi}), \Psi_{II} (\alpha, \vec{\phi}) \rangle =~~~~~~~~~~~~~~~~~~~~~~~~~~
\nonumber\\
i \int d^N \phi\Bigl(\Psi^*_{I} (\alpha, \vec{\phi}) \dot{\Psi}_{II} (\alpha, \vec{\phi}) - \Psi_{II} (\alpha, \vec{\phi}) \dot{\Psi}^*_{I} (\alpha, \vec{\phi}) \Bigr) \label{inn prod}
\end{eqnarray}
on each space-like hypersurface $\alpha =$  constant. Here and hereafter dots denote the derivatives with respect to the intrinsic time $\alpha$.
One may be tempted to quantize
field with respect to the inner product (\ref{inn prod}) as
\begin{eqnarray}
\hat{\Psi} (\alpha, \vec{\phi}) = \int \frac{d^N \phi}{(2 \pi)^{N/2}} \bigl(
e^{ i \vec{p} \cdot \vec{\phi} } \psi_{\vec{p}} (\alpha) \hat{a}_{\vec{p}} + {\rm H.~C.} \bigr),
\label{sec quan}
\end{eqnarray}
where $\psi_{\vec{p}} (\alpha)$ is a positive frequency solution with respect to $\alpha$,
${\rm H.~C.}$ denotes the Hermitian conjugate, and $\hat{a}_{\vec{p}}$ and its
Hermitian conjugate $\hat{a}^{\dagger}_{\vec{p}}$ are Schr\"{o}dinger operators.
The quantization (\ref{sec quan}) holds  as far as $\psi^*_{\vec{p}} (\alpha)$ is a negative frequency solution, which is not necessarily true for the closed universe as will be shown in Sec. \ref{sec-IV}.
We may use the annihilation and the creation operators
to construct the Hilbert space consisting of the vacuum of
no-universe (nothingness) and the one-universe and the multi-universes
\begin{eqnarray}
{\bf C} \oplus {\bf H} \oplus ({\bf H} \otimes {\bf H}) \oplus ({\bf H} \otimes {\bf H} \otimes {\bf H}) \oplus \cdots,
\end{eqnarray}
where ${\bf C}$ denotes the vacuum state
\begin{eqnarray}
\hat{a}_{\vec{p}} \vert 0 \rangle = 0 \quad ({\rm for\, all\,} \vec{p}),
\end{eqnarray}
and ${\bf H}$ denotes the single universe
\begin{eqnarray}
\hat{a}^{\dagger}_{\vec{p}} \vert 0 \rangle \quad ({\rm for\, any\,} \vec{p})),
\end{eqnarray}
and ${\bf H} \otimes {\bf H}$ denotes double universes
\begin{eqnarray}
\hat{a}^{\dagger}_{\vec{p}} \hat{a}^{\dagger}_{\vec{p}'} \vert 0 \rangle \quad ({\rm for\, any\,} \vec{p}\, {\rm and}\, \vec{p}'),
\end{eqnarray}
and so on.

As will be shown in the following sections, the third quantization of the super-action may shed light on
interpreting the quantum states, in particular, of the closed universe.

\section{Third Quantized Formulation} \label{sec-III}

In the third quantized formulation, the WDW equation can be obtained from the super-action in the superspace
\begin{eqnarray}
{\cal S} = \frac{1}{2} \int d \alpha d^N {\bf \phi} \Bigl[ \Bigl(\frac{\partial \Psi}{\partial \alpha} \Bigr)^2
- \Bigl(\nabla_{\vec{\phi}} \Psi \Bigr)^2 - {\cal M}^2 (\alpha) \Psi^2\Bigr]. \label{3-act}
\end{eqnarray}
The variation $\delta {\cal S}/\delta \Psi$ leads to the WDW equation (\ref{wdw eq}).
We expand the wave function by the Fourier mode as
\begin{eqnarray}
\Psi (\alpha, \phi) =  \int \frac{d^N p}{(2 \pi)^{N/2}} \psi_{\vec{p}} (\alpha) e^{i \vec{p} \cdot \vec{\phi}}.
\end{eqnarray}
Redefining the field as
\begin{eqnarray}
\psi_{(+)  \vec{p}} = \frac{1}{2} (\psi_{\vec{p}} + \psi_{- \vec{p}}),
\quad \psi_{(-) \vec{p}} = \frac{1}{2i} (\psi_{\vec{p}} - \psi_{-\vec{p}}),
\end{eqnarray}
we obtain
\begin{eqnarray}
\int d^N {\bf \phi} \Psi^2 (\alpha, \vec{\phi}) &=& \int d^N p \bigl(\psi^2_{(+)\vec{p}}
+ \psi^2_{(-)\vec{p}} \bigr), \nonumber\\
\int d^N {\bf \phi} \dot{\Psi}^2 (\alpha, \vec{\phi}) &=& \int d^N p \bigl(\dot{\psi}^2_{(+)\vec{p}}
+ \dot{\psi}^2_{(-)\vec{p}} \bigr).
\end{eqnarray}
Therefore, the Hamiltonian in the third quantization takes the form
\begin{eqnarray}
{\cal H} (\alpha) = \frac{1}{2} \sum_{\pm} \int d^N p \bigl[ \dot{\pi}^2_{(\pm) \vec{p}}
+ \omega_p^2 (\alpha) \psi^2_{(\pm) \vec{p}} \bigr], \label{super ham}
\end{eqnarray}
where $\pi_{(\pm) \vec{p}} = \dot{\psi}_{(\pm) \vec{p}}$
and the time-dependent frequency is
\begin{eqnarray}
\omega_p^2 (\alpha) = \vec{p}^2 - k e^{4 \alpha}. \label{freq}
\end{eqnarray}
The Hamiltonian (\ref{super ham}) is an infinite sum of time-dependent oscillators.
For the open $(k = -1)$ and the flat $(k = 0)$ universes, the frequencies always take real values, so
the Hamiltonian (\ref{super ham}) describes ordinary time-dependent oscillators.
However, for the closed universe $(k = 1)$ the squared frequencies become negative when $e^{4 \alpha } > p^2$,
and thus the universe has a maximum size $e^{\alpha_0} = a_0 = \sqrt{p}$ at the classical level,
beyond of which it becomes tachyonic at the quantum level.

\subsection{Quantum States of Universes}

The quantum law for the universe is the intrinsic time-dependent
Schr\"{o}dinger equation
\begin{eqnarray}
i \frac{\partial}{\partial \alpha} \Psi (\alpha) = \hat{\cal H} (\alpha) \Psi (\alpha).
\end{eqnarray}
As the Hamiltonian is decoupled, the wave function is the product of the wave function for
each Fourier mode:
\begin{eqnarray}
\Psi (\alpha) = \prod_{(\pm) \vec{p}} \Psi_{(\pm) \vec{p}} (\alpha), \label{wdw st}
\end{eqnarray}
where each Fourier-mode independently obeys the Sch\"{o}dinger equation
\begin{eqnarray}
i \frac{\partial}{\partial \alpha} \Psi_{(\pm) \vec{p}} (\alpha) = \hat{\cal H}_{(\pm) \vec{p}}
(\alpha) \Psi_{(\pm) \vec{p}} (\alpha). \label{osc}
\end{eqnarray}
The time-dependent Schr\"{o}dinger equation has the well-known invariant discovered by Lewis and Riesenfeld, whose eigenstates provide the exact quantum states up to time-dependent phase factors \cite{lewis-riesenfeld}.
Here we shall use a pair of invariants of the form \cite{MMT,kim-kim99,kim-page01}
\begin{eqnarray}
\hat{a}_{(\pm) \vec{p}} (\alpha) &=& i \bigl[ \bar{u}_{\vec{p}} (\alpha) \hat{\pi}_{(\pm) p} - \dot{\bar{u}}_{\vec{p}} (\alpha) \hat{\psi}_{(\pm) \vec{p}} \bigr], \nonumber\\
\hat{a}^{\dagger}_{(\pm) \vec{p}} (\alpha) &=& -i \bigl[ u_{\vec{p}} (\alpha) \hat{\pi}_{(\pm) p} - \dot{u}_{\vec{p}} (\alpha) \hat{\psi}_{(\pm) \vec{p}} \bigr], \label{inv pair}
\end{eqnarray}
for two independent solutions $u_{\vec{p}} (\alpha)$ and $\bar{u}_{\vec{p}} (\alpha)$
of eq. (\ref{grav eq}), which
satisfy the Liouville-von Neumann equation
\begin{eqnarray}
i \frac{\partial \hat{a}_{(\pm) \vec{p}} (\alpha)}{ \partial \alpha}
+ \bigl[
\hat{a}_{(\pm) \vec{p}} (\alpha), \hat{\cal H}_{(\pm) \vec{p}} (\alpha) \bigr] = 0.
\end{eqnarray}
The Wronskian condition from the quantization rule of the field (\ref{inn prod})
\begin{eqnarray}
{\rm Wr} [u_{\vec{p}} (\alpha),\bar{u}_{\vec{p}} (\alpha)] = - i. \label{wr con}
\end{eqnarray}
makes $\hat{a}_{(\pm) \vec{p}} (\alpha)$ and $\hat{a}^{\dagger}_{(\pm)\vec{p}} (\alpha)$
the time-dependent annihilation and creation operators with the equal-time commutation relation
\begin{eqnarray}
[\hat{a}_{(\pm) \vec{p}} (\alpha), \hat{a}^{\dagger}_{(\pm)\vec{p}'} (\alpha)] = \delta_{\vec{p} \vec{p}'}.
\end{eqnarray}

The quantum states for each oscillator can be constructed from the creation and
the annihilation operators (\ref{inv pair}).
In the ket-vector representation, the ground state is
\begin{eqnarray}
\hat{a}_{(\pm) \vec{p}} (\alpha) \vert 0_{(\pm) \vec{p}}, \alpha \rangle = 0,
\end{eqnarray}
and the number states are
\begin{eqnarray}
\vert n_{(\pm) \vec{p}}, \alpha \rangle = \frac{\bigl( \hat{a}^{\dagger }_{(\pm) \vec{p}} (\alpha) \bigr)^{n_{(\pm) \vec{p}}} }{\sqrt{ n_{(\pm) \vec{p}}! } } \vert 0_{(\pm) \vec{p}}, \alpha \rangle.
\end{eqnarray}
The vacuum for the WDW equation of the form (\ref{wdw st}) is given by
\begin{eqnarray}
\vert 0, \alpha \rangle = \prod_{(\pm) \vec{p}} \vert 0_{(\pm) \vec{p}}, \alpha \rangle.
\end{eqnarray}
Another interesting state is the coherent state
\begin{eqnarray}
\hat{a}_{(\pm) \vec{p}} (\alpha) \vert z_{(\pm) \vec{p}}, \alpha \rangle = z_{(\pm) \vec{p}} \vert z_{(\pm) \vec{p}}, \alpha \rangle,
\end{eqnarray}
for a complex constant $z_{(\pm) \vec{p}}$. In the case of  $\bar{u}_p =  u^*_p$
the coordinate representation of the bra-vector is the complex conjugate
of that of the ket-vector.

However, in the case of $\bar{u}_p \neq  u^*_p$, the bra-vector representation for the
dual-ground state is
\begin{eqnarray}
\langle 0_{(\pm) \vec{p}}, \alpha \vert \hat{a}^{\dagger}_{(\pm)\vec{p}} (\alpha) = 0,
\end{eqnarray}
and for the dual-coherent state is
\begin{eqnarray}
\langle  \bar{z}_{(\pm) \vec{p}}, \alpha \vert \hat{a}^{\dagger}_{(\pm) \vec{p}} (\alpha) = \bar{z}_{(\pm) \vec{p}} \langle  \bar{z}_{(\pm) \vec{p}}, \alpha \vert,
\end{eqnarray}
for another constant complex $\bar{z}_{(\pm) \vec{p}}$.
The coordinate representation is for the ket-vector
\begin{eqnarray}
\Psi (\psi_{(\pm) \vec{p}}, \alpha) &=&
 \langle \psi_{(\pm) \vec{p}} \vert 0, \alpha \rangle \nonumber\\&=& \frac{1}{\sqrt{\sqrt{\pi} \bar{u}_p} } \exp \Bigl[i \frac{\dot{\bar{u}}_p}{\bar{u}_p} \psi^2_{(\pm) p} \Bigr],
\label{gaus ket}
\end{eqnarray}
and for the bra-vector
\begin{eqnarray}
\bar{\Psi}(\psi_{(\pm) \vec{p}}, \alpha) &=& \langle 0, \alpha \vert \psi_{(\pm) \vec{p}} \rangle
\nonumber\\ &=& \frac{1}{\sqrt{\sqrt{\pi} u_p} } \exp \Bigl[-i \frac{\dot{u}_p}{u_p} \psi^2_{(\pm) p} \Bigr]. \label{gaus bra}
\end{eqnarray}
Note that the orthonormality of the bra- and ket-vectors, $\langle 0, \alpha \vert 0, \alpha \rangle = 1$,
is realized as
\begin{eqnarray}
\int d\psi_{(\pm) \vec{p}} \bar{\Psi} (\psi_{(\pm) \vec{p}}, \alpha)
\Psi (\psi_{(\pm) \vec{p}}, \alpha) = 1,
\end{eqnarray}
due to eq. (\ref{wr con}).
Then the Fock space of ket-vectors and bra-vectors leads to the propagator
for the WDW equation of the form
\begin{eqnarray}
\prod_{(\pm) \vec{p}} \Bigl(\sum_{n_{(\pm) \vec{p}}=0}^{\infty}  \vert n_{(\pm) \vec{p}}, \alpha \rangle \langle n_{(\pm) \vec{p}}, \alpha_0 \vert \Bigr).
\end{eqnarray}

Using the Wronskian condition (\ref{wr con}), the field and the momentum operators are expressed as
\begin{eqnarray}
\hat{\psi}_{(\pm) p} &=& u_{\vec{p}} (\alpha) \hat{a}_{(\pm) \vec{p}} (\alpha)+
\bar{u}_{\vec{p}} (\alpha) \hat{a}^{\dagger}_{(\pm) \vec{p}} (\alpha), \nonumber\\
\hat{\pi}_{(\pm) p} &=& \dot{u}_{\vec{p}} (\alpha) \hat{a}_{(\pm) \vec{p}} (\alpha) +
\dot{\bar{u}}_{\vec{p}} (\alpha) \hat{a}^{\dagger}_{(\pm) \vec{p}} (\alpha). \label{disp 1}
\end{eqnarray}
The third quantized field is given in terms of pairs of the invariant operators as
\begin{eqnarray}
\hat{\Psi} (\alpha, \vec{\phi}) &=& \int \frac{d^N \phi}{(2 \pi)^{N/2}} \Bigl[
e^{ i \vec{p} \cdot \vec{\phi} } u_{\vec{p}} \bigl( \hat{a}_{(+) \vec{p}}
+ i \hat{a}^{\dagger}_{(+) \vec{p}} \bigr) \nonumber\\
&&+ e^{-i \vec{p} \cdot \vec{\phi} } u_{\vec{p}} \bigl(\hat{a}_{(-) -\vec{p}}
+ i \hat{a}^{\dagger}_{(-) -\vec{p}} \bigr) \Bigr].
\end{eqnarray}
One can show that
\begin{eqnarray}
\langle 0, \alpha \vert \hat{\psi}_{(\pm) \vec{p}} \vert 0, \alpha \rangle = \langle 0, \alpha \vert \hat{\pi}_{(\pm) \vec{p}} \vert 0, \alpha \rangle = 0 \label{disp 1}
\end{eqnarray}
and
\begin{eqnarray}
\langle 0, \alpha \vert \hat{\psi}_{(\pm) \vec{p}}^2 \vert 0, \alpha \rangle &=& \bar{u}_{\vec{p}}
(\alpha) u_{\vec{p}} (\alpha), \nonumber\\
\langle 0, \alpha \vert \hat{\pi}_{(\pm) \vec{p}}^2 \vert 0, \alpha \rangle &=& \dot{\bar{u}}_{\vec{p}} (\alpha) \dot{u}_{\vec{p}} (\alpha). \label{disp 2}
\end{eqnarray}
Further, the coherent state follows a classical trajectory
\begin{eqnarray}
\langle  \bar{z}_{(\pm) \vec{p}}, \alpha \vert \hat{\psi}_{(\pm) \vec{p}} \vert z_{(\pm) \vec{p}}, \alpha \rangle = \nonumber\\
 z_{(\pm) \vec{p}} u_{(\pm) \vec{p}} (\alpha) + \bar{z}_{(\pm) \vec{p}}
\bar{u}_{(\pm) \vec{p}} (\alpha),
\end{eqnarray}
provided that $\bar{z}_{(\pm) \vec{p}}
\bar{u}_{(\pm) \vec{p}} = (z_{(\pm) \vec{p}} u_{(\pm) \vec{p}})^* $.

\section{Open and Closed Universes} \label{sec-IV}

The WDW equation for a spatially flat FRW universe is trivial as the Klein-Gordon equation
in the Minkowski spacetime and will not be considered further in this paper. The open universe becomes
an infinite system of intrinsic time-dependent oscillators while the closed universe
shows an interesting feature of tachyonic behavior for the super-Planckian size (sub-Planckian energy) universes for given momenta for massless fields. However, both universes become ultra-relativistic
at the sub-Planckian scale (super-Planckian energy) and behave as a free field with the given
momenta. We shall separately treat the open and the closed universes below.

\subsection{Open Universe}

The sub-Planckian size universe $(e^{\alpha} \ll \sqrt{p})$ belongs to a ultra-relativistic  regime dominated by
the kinetic energy of massless fields and has the solutions
\begin{eqnarray}
u_{p} (\alpha) &=& \frac{2^{-ip}}{\sqrt{2p}} \Gamma (1- ip/2) J_{-i p/2} \Bigl( \frac{e^{2 \alpha}}{2}  \Bigr), \nonumber\\
\bar{u}_{p} (\alpha) &=& u^*_{p} (\alpha). \label{k=-1 small}
\end{eqnarray}
Then the solutions are asymptotically
$u_p = e^{- i p \alpha}/\sqrt{2p}$
and $\bar{u}_p = e^{i p \alpha}/\sqrt{2p}$, which correspond to the positive and
negative frequency solutions with
respect to $i \partial/\partial \alpha$, respectively. The annihilation and the creation operators
in eq. (\ref{inv pair}) with (\ref{k=-1 small}) allow us to construct the
Gaussian wave function, the number-state wave functions and the coherent state.
From the dispersion relations (\ref{disp 1}) and (\ref{disp 2}) we find that
\begin{eqnarray}
\Delta \psi_{(\pm) p}^2 = \frac{1}{2p}, \quad
\Delta \pi_{(\pm) p}^2 = \frac{p}{2}.
\end{eqnarray}
Thus the Gaussian wave function has the standard dispersion for a simple harmonic oscillator
and keeps the minimum uncertainty.

On the other hand, the super-Planckian size universe $(e^{\alpha} \gg \sqrt{p})$
has the positive and negative frequency solutions
\begin{eqnarray}
v_{p} (\alpha) &=& \sqrt{\frac{\pi}{8}} e^{-i \frac{\pi}{4}} e^{\frac{p\pi}{4}}
H_{i p/2}^{(2)} \Bigl( \frac{e^{2 \alpha}}{2} \Bigr), \nonumber\\
\bar{v}_{p} (\alpha) &=& \sqrt{\frac{\pi}{8}} e^{i \frac{\pi}{4}} e^{- \frac{p\pi}{4}} H_{ip/2}^{(1)} \Bigl( \frac{e^{2 \alpha}}{2} \Bigr). \label{k=-1 large}
\end{eqnarray}
Here $ v_{p}$ for $u_{p}$ and $ \bar{v}_{p}$ for $\bar{u}_{p}$ in eq. (\ref{inv pair})
lead to another pairs of invariant operators
\begin{eqnarray}
\hat{b}_{(\pm) \vec{p}} (\alpha) &=& i \bigl[ \bar{v}_{\vec{p}} (\alpha) \hat{\pi}_{(\pm) p} - \dot{\bar{v}}_{\vec{p}} (\alpha) \hat{\psi}_{(\pm) \vec{p}} \bigr], \nonumber\\
\hat{b}^{\dagger}_{(\pm) \vec{p}} (\alpha) &=& -i \bigl[ v_{\vec{p}} (\alpha) \hat{\pi}_{(\pm) p} - \dot{v}_{\vec{p}} (\alpha) \hat{\psi}_{(\pm) \vec{p}} \bigr]. \label{inv pair2}
\end{eqnarray}
The solution (\ref{k=-1 large}) can be analytically continued to
the sub-Planckian size universe \cite{table}
\begin{eqnarray}
v_p (\alpha) = \mu_{p} u_p (\alpha) + \nu_p \bar{u}_p (\alpha), \label{bog rel}
\end{eqnarray}
where
\begin{eqnarray}
\mu_p &=& \sqrt{\frac{p \pi}{4}}  \frac{2^{ip} e^{-i \frac{\pi}{4}} e^{\frac{p\pi}{4}}}{ \Gamma (1 - i p/2) \sinh (p \pi/2)}, \nonumber\\ \nu_p &=& - \sqrt{\frac{p \pi}{4}} \frac{2^{- ip} e^{-i \frac{\pi}{4}} e^{- \frac{p\pi}{4}}}{ \Gamma (1 + i p/2) \sinh (p \pi/2)}.
\end{eqnarray}
Note that the Bogoliubv relation $|\mu_p|^2 - |\nu_p|^2 = 1$ holds.
The linear relation (\ref{bog rel}) leads to the Bogoliubov transformation
\begin{eqnarray}
\hat{b}_{(\pm) p} (\alpha) = \mu_p^* \hat{a}_{(\pm) p} (\alpha) - \nu_p^* \hat{a}_{(\pm) p}^{\dagger} (\alpha). \label{bog tran}
\end{eqnarray}
Now the dispersion relations
\begin{eqnarray}
\Delta \psi_{(\pm) p}^2 = \frac{1}{2e^{2 \alpha}}, \quad
\Delta \pi_{(\pm) p}^2 = \frac{e^{2 \alpha}}{2},
\end{eqnarray}
show that the Gaussian wave packet is sharply peaked around $\psi_{(\pm) p} = 0$ at the cost
of the widely broadened dispersion in the momentum space. However, the
uncertainty keeps the minimum value at the leading order.

In the third quantization, the super-Planckian size universe has
the Planckian distribution of the sub-Planckian size universes
\begin{eqnarray}
N_p = \frac{1}{e^{p \pi} -1}.
\end{eqnarray}

\subsection{Closed Universe}

The quantization of the closed universe is more intriguing than the open or flat universe since
the frequency for each oscillator changes the sign after $e^{\alpha_0} = \sqrt{p}$ and
thus becomes unstable. The inverted quantum oscillators occur in the slow-rolling inflation model
\cite{guth-pi85} and in the second order phase transition \cite{kim-lee00}.

Now the solutions for sub-Planckian size universes are
\begin{eqnarray}
u_{p} (\alpha) &=& \frac{2^{-ip}}{\sqrt{2p}} e^{-\frac{p\pi}{4}} \Gamma (1- ip/2)  J_{-i p/2} \Bigl( \frac{ie^{2 \alpha}}{2}\Bigr), \nonumber\\
\bar{u}_{p} (\alpha) &=& \frac{2^{ip}}{\sqrt{2p}} e^{\frac{p\pi}{4}} \Gamma (1+ ip/2)  J_{i p/2} \Bigl( \frac{ie^{2 \alpha}}{2} \Bigr). \label{k=1 small}
\end{eqnarray}
It should be noted that $\bar{u}_{p}$ is not the complex conjugate of $u_{p}$ but the replacement of
$p$ by $-p$. Thus the Gaussian wave function (\ref{gaus bra}) for the bra-vector is not the
complex conjugate of (\ref{gaus ket}) for the ket-vector. These two Gaussian wave packets
form an orthonormal pair. With respect to these two Gaussian wave packets the dispersion relations
take the asymptotic values
\begin{eqnarray}
\Delta \psi_{(\pm) p}^2 = \frac{1}{2p}, \quad
\Delta \pi_{(\pm) p}^2 = \frac{p}{2}.
\end{eqnarray}
In fact, the asymptotic form of eq. (\ref{k=1 small}) is the same as that
of eq. (\ref{k=-1 small}) for the open universe up to a constant.

On the other hand, for the super-Planckian size universe, the invariant operators (\ref{inv pair}) may constructed from
\begin{eqnarray}
v_p (\alpha) &=& \frac{\sqrt{\pi} S}{2}  I_{-ip/2} \Bigl( \frac{ie^{2 \alpha}}{2} \Bigr)
+ i \frac{\sqrt{\pi}}{2 S} K_{-ip/2} \Bigl( \frac{ie^{2 \alpha}}{2} \Bigr),  \nonumber\\
\bar{v}_p (\alpha) &=& \frac{\sqrt{\pi} S}{2}  I_{-ip/2} \Bigl( \frac{i}{2} e^{2 \alpha}\Bigr)
- i \frac{\sqrt{\pi}}{2 S} K_{-ip/2} \Bigl( \frac{i}{2} e^{2 \alpha}\Bigr). \nonumber\\
\label{k=1 large}
\end{eqnarray}
Here $S$ is a real constant to be determined later.
The solutions (\ref{k=1 large}) have the asymptotic forms
\begin{eqnarray}
v_p (\alpha) &=& \frac{1}{\sqrt{4 e^{2 \alpha}}} \Bigl( S e^{\frac{1}{2} e^{2 \alpha}}
+ \frac{i}{S} e^{- \frac{1}{2} e^{2 \alpha}} \Bigr), \nonumber\\
\bar{v}_p (\alpha) &=& \frac{1}{\sqrt{4 e^{2 \alpha}}} \Bigl( S e^{\frac{1}{2} e^{2 \alpha}}
- \frac{i}{S} e^{- \frac{1}{2} e^{2 \alpha}} \Bigr),
\label{k=1 asym}
\end{eqnarray}
and, therefore, the dispersion relations are given by
\begin{eqnarray}
\Delta \psi_{(\pm) p}^2 &=& \frac{1}{4 e^{2 \alpha}} \Bigl( S^2 e^{e^{2 \alpha}}
+ \frac{1}{S^2} e^{-e^{2 \alpha}} \Bigr), \nonumber\\
\Delta \pi_{(\pm) p}^2 &=& \frac{e^{2 \alpha}}{4 } \Bigl( S^2 e^{e^{2 \alpha}}
+ \frac{1}{S^2} e^{- e^{2 \alpha}} \Bigr).
\end{eqnarray}
The googolplex growing dispersion in the position implies that any measurement
of localization of the universe may fail.

Using the relations of Bessel functions \cite{table} in a Riemann sheet $- \pi < z \leq \pi/2$
\begin{eqnarray}
I_{\nu} (z) &=& e^{- \frac{\pi}{2} \nu i} J_{\nu} \bigl(i z \bigr), \nonumber\\
K_{\nu} (z) &=& i \frac{\pi}{2} e^{\frac{\pi}{2} \nu i} H^{(1)}_{\nu} \bigl(i z \bigr), \nonumber\\
H^{(1)}_{\nu} (z) &=& \frac{i e^{-i \nu \pi}}{\sin (\nu \pi)}J_{\nu} (z)
- \frac{i}{\sin(\nu \pi)} J_{\nu} (z),
\end{eqnarray}
we may find the relation
\begin{eqnarray}
v_p (\alpha) = \mu_p u_p (\alpha) + \nu_p \bar{u}_p (\alpha).
\end{eqnarray}
Here the Bogoliubov coefficients are given by
\begin{eqnarray}
\mu_p &=& \sqrt{\frac{p \pi}{4} } \Bigl[ 2 e^{- \frac{p \pi}{4}} +
\Bigl(S  e^{- \frac{p \pi}{4}} + \frac{ e^{\frac{p \pi}{4}} }{S} \Bigr)
\frac{ e^{- \frac{p \pi}{2}} }{\sinh (\frac{p \pi}{2}) } \Bigr] \nonumber\\
&& \times
\frac{2^{ip} e^{\frac{p\pi}{4}} }{ \Gamma (1 - i p/2)}, \nonumber\\ \nu_p &=& -
 \sqrt{\frac{p \pi}{4} }
\Bigl(S  e^{- \frac{p \pi}{4}} + \frac{ e^{\frac{p \pi}{4}} }{S} \Bigr)
\frac{ e^{- \frac{p \pi}{2}} }{\sinh (\frac{p \pi}{2}) } \nonumber\\
&& \times
\frac{2^{-ip} e^{- \frac{p\pi}{4}} }{ \Gamma (1 + i p/2)}.
\end{eqnarray}
The Bogoliubov relation $|\mu_p|^2 - |\nu_p|^2 = 1$ holds, provided $S$ satisfies
\begin{eqnarray}
S^2 +  e^{\frac{p \pi}{4}} \Bigl( e^{\frac{p \pi}{2}}+ e^{- \frac{p \pi}{2}}
+ \sqrt{e^{p \pi}-1} \Bigr) S + e^{\frac{p \pi}{2}} = 0.
\end{eqnarray}

Then the number of sub-Planckian size universes contained in the super-Planckian size universe
is
\begin{eqnarray}
N_p = \frac{1}{e^{p \pi} -1}
\Bigl[e^{\frac{p \pi}{2}} \Bigl( e^{\frac{p \pi}{2}}+ e^{- \frac{p \pi}{2}}
+ \sqrt{e^{p \pi}-1} \Bigr)^2 \Bigr].
\end{eqnarray}
The terms in the square bracket is an amplification factor for production of
super-Planckian size universes.

\section{Conclusion}

Quantum cosmology may be a consistent framework for studying quantum fluctuations of spacetime and matters. In particular, the WDW equation has one-to-one correspondence with the
variables for classical cosmology. Further, the wave packet is peaked along the quasi-classical trajectory
in the superspace. The large disparity between the Planck mass and
the matter fields may separate gravity
from matter fields a la the Born-Oppenheimer idea of separating
heavy particles from light particles and the Born-pilot theory of writing quantum law as the Hamilton-Jacobi equation
 together with the continuity equation for the amplitude. This leads to the semiclassical gravity, in which gravity becomes classical while matter fields still obey the unitary quantum field theory.
This quantum-to-(semi)classical transition may resolve the back-reaction problem of gravitational and matter fluctuations at the one-loop level. At least quantum cosmological models
with a few gravitational degrees of freedom
and matter fields such as inflatons could avoid fundamental questions on quantum gravity such as
renormalization, unitarity, etc.

In this paper we have studied quantum cosmology for the FRW universe with $N$-massless scalar fields.
The massless fields may have an analog of radiation in the FRW universe.
Compact spaces with constant curvature in some multidimensional
cosmological models play the same role of massless fields.
The quantum FRW cosmological model with massless fields is exactly
solvable as shown in this paper. Further, the quantum FRW cosmological model
for the closed universe exhibits an intriguing feature
that the forbidden region beyond the classical size suffers from an instability
for quantum universes.
It has an analog of second order phase transitions quenched by some external agents.
We have constructed the quantum states for the universe beyond the classical region. The dispersion relation of the position of the field
in the third quantization grows as googolplex, which makes any measurement of the
size of the universe meaningless far beyond the classical regime. Further, the googolplex growing
uncertainty also suggests that quantum universes in the classically forbidden region behave as classical.

Keeping in mind the quantum-to-(semi)classical transition, it would be interesting to consider
the quantum cosmological models with single-field inflatons consistent with the current CMB data such as
the seven-year WMAP and SPT, etc. For instance, the massive scalar field is still
viable from the observational data. As shown in refs. \cite{kim92,mostafazadeh98}
the WDW equation could not be separated by mode-by-mode; instead,
the gravitational field part expanded by
the eigenfunctions of the massive scalar field takes a coupled vector equation.
In fact, the instantaneous
eigenfunctions change as a function of the intrinsic time, whose evolution is governed by
the coupling matrix \cite{kim92}. The third quantized FRW universe with a massive scalar field is analogous to the Klein-Gordon equation in a homogeneous and time-dependent magnetic field, whose transverse motion corresponds to the scalar field in the WDW equation. Thus the third quantized formulation of the massive scalar field cosmology is an infinite number of coupled time-dependent
oscillators, whose study will be advanced in future publication.

\section{Acknowledgments}
The author would like to thank Pauchy W-Y. Hwang and Guey-Lin Lin for the warmest hospitality
at the 9th Cosmology and Particle Astrophysics (CosPA) Symposium, National Taiwan University and National Chiao Tung University and also thank Francois Bouchet and Je-An Gu for bringing attention to ref. \cite{SPT}, which has motivated the revival of the massive field quantum cosmology in ref. \cite{kim92}.
He also thanks the Yukawa Institute for Theoretical Physics at Kyoto University,
where this work was completed during the Long-term Workshop YITP-T-12-03 on "Gravity and Cosmology 2012".
This work was supported in part by Basic Science Research Program through
the National Research Foundation of Korea (NRF) funded by the Ministry of Education, Science and
Technology (22012R1A1B3002852).







\end{document}